\documentclass[12pt]{iopart}
\usepackage[utf8]{inputenc}
\usepackage{graphicx}

\usepackage{graphicx}
\usepackage{subfig}
\usepackage{braket}
\usepackage{bbold}
\usepackage{cite}

\begin{document}

\title{Evolutionary algorithm to design high-cooperativity optical cavities}

\date{January 2022}

\author{D. V. Karpov, P. Horak}

\address{Optoelectronics Research Centre, University of Southampton, Southampton SO17 1BJ, United Kingdom}
\ead{d.karpov@soton.ac.uk}

\begin{abstract}
Using an evolutionary algorithm combined with a gradient descent method we design optical cavities with significantly enhanced strong coupling rates between cavity photons and a single quantum emitter. Our approach allows us to find specially designed non-spherical mirrors which lead to high-finesse cavity eigenmodes with large field enhancement at the center of the cavity. The method is based on adding consecutive perturbations to an initial spherical mirror shape using the gradient descent method for optimization. We present mirror profiles suitable for fabrication which demonstrate higher cavity cooperativity than any spherical cavity of the same size. Finally, we demonstrate numerically how such a cavity enhances the operation frequency and purity of coupling a Ca$^+$ ion to an optical fiber photon.
\end{abstract}

\maketitle


\section{Introduction}

Strong coupling of a quantum emitter, e.g., an ion, atom, NV-center, or quantum dot, to an optical mode of a resonator and long cavity photon lifetime are essential in numerous applications of quantum optics for fundamental research and for practical quantum technology. Promising systems to fulfill these requirements are fiber-optic microcavities \cite{Pellizzari1995,Cirac1997,Kimble2008,Monroe2013}, ion beam etched dielectric resonators \cite{fib}, or micro-assembled structures \cite{revResonators}.

Strong coupling between emitter and cavity photon can be realized by a small cavity volume and therefore by a very short optical cavity. However, for many realistic quantum setups the cavity mirrors cannot be placed too close to each other due to technical difficulties: for trapped ion systems, short cavities lead to electrical charging of the dielectric mirrors and to distortion of the radio frequency ion trapping fields \cite{comp}; for neutral atoms, short cavity lengths are limited by the requirement for delivery of atoms into the cavity and for optical side access \cite{Harlander2010,Podoliak2016} for cooling and trapping. Thus, optical cavities utilized in quantum optical device applications need to combine a strong coupling rate with low losses while keeping the mirrors sufficiently apart. 

One way to achieve strong coupling is to operate the cavity in a (near-) concentric configuration \cite{yariv}. This minimizes the optical mode field waist at the cavity center, thus maximizing emitter-photon coupling, but at the cost of increased clipping losses because of a large mode field diameter on the mirrors and extreme sensitivity of the cavity to alignment accuracy.

Another method to increase the field amplitude in the center of the cavity is the creation of some interference pattern by modulating the mirror profile \cite{paper1}. We assume that we are not limited by spherical cavities, i.e., we can create an arbitrary mirror shape using, for example, focused ion beam milling or laser ablation  as discussed in more detail in Sec.\ \ref{sec:conclusion}. Here, we numerically explore modulated spherical profiles of the cavity mirrors which give rise to highly localized cavity modes while at the same time keeping losses low. With this approach we find a manifold of mirror profiles which can provide a lower loss rate than a concentric cavity but avoid the mode instability of the concentric cavity configuration.

The goal of our work here is to present a procedure based on an evolutionary algorithm which allows us to design mirror shapes for optical cavities with higher coupling and cooperativity compared to spherical cavity mirrors of the same size. We then apply the method to optimize a fiber-tip cavity by also taking into account out-coupling of a cavity photon into an optical fiber, This yields a (slightly) larger overall transfer efficiency of an ion excitation to a fiber photon, but significantly increases the operating repetition frequency of such a device.

Evolutionary or genetic algorithms are used for various applications mostly for inverse design problems \cite{evo1,evo2} and have also been applied in the field of modern optics \cite{evo3,evo4,evo5}. In forward design of optical cavities, for given geometry and boundary conditions the corresponding cavity optical field can be calculated by highly developed and well known numerical, semi-analytical or analytical methods. However, it is not often possible to find parameters to create a target optical field distribution using forward design methods. Here we demonstrate a method which is based on consecutive ``mutations'' of an initial spherical mirror shape while calculating the mode spectrum and conducting gradient descent to optimize the mutations. The method allows us to find mirror profiles which provide significant enhancement of a target parameter, such as the cavity cooperativity.

This paper is organized as follows. First, in Sec.\ \ref{sec:statement} we describe the problem under investigation, our  motivation, theoretical model and optimization parameters.  In Sec.\ \ref{sec:algo} we describe the algorithm and the role of the gradient descent method in it. In Sec.\ \ref{sec:results} we present the results of our simulations, in particular the cavity mode topology, strong coupling rate and cooperativity enhancement factor. In Sec.\ \ref{sec:lambda} we apply our method for coupling a Ca$^+$ ion in a $\Lambda$-scheme via a fiber-tip cavity to an optical fiber photon and demonstrate the significantly faster time evolution of the improved system. Finally, we briefly discuss fabrication opportunities to realize such mirrors experimentally and conclude in Sec.\ \ref{sec:conclusion}. 



\section{Problem statement}\label{sec:statement}

The coherent coupling between a quantum emitter, such as a quantum dot, ion or cold atom, located at a coordinate $\textbf{r}$ in an cavity with a dimensionless optical field mode $\psi(\textbf{r})$ is characterized by the strong coupling rate \cite{Vucovich}
\begin{equation}
g_0(\textbf{r})=\sqrt{\frac{3\lambda^2c\Gamma}{4\pi V_{\psi} }}\psi(\textbf{r}),\
\psi(\textbf{r}) = \frac{E(\textbf{r})}{|E(\textbf{r}_m)|}
\label{eq:coupl1}
\end{equation}
where $E(\textbf{r})$, $\Gamma$, $\lambda$, $\textbf{r}_m$ are electric field of the cavity mode, the spontaneous decay rate of the emitter 
$\Gamma= \frac{\omega^3 \mu^2}{3\pi\varepsilon_{0}\hbar c^3} $ 
(where $\omega$ is the transition angular frequency and $\mu$ is the electric dipole moment), its transition wavelength, and the maximum electric field intensity point, respectively. Here we assume that the emitter dipole moment and the electric field polarization vectors are aligned. The cavity mode volume $V_{\psi}$ is given by
\begin{equation}
V_{\psi} = \int_{V_{cavity}} |\psi(\textbf{r})|^2 dV.
\label{eq:volume}
\end{equation}

If the emitter-cavity strong coupling rate $g_0$ is larger than the strength of any incoherent processes, i.e., than the decay rate $\Gamma$ and the loss rate $\kappa$ of the cavity field, the cavity operates in the strong coupling regime. In this case the cooperativity parameter $C_0$ at the field maximum,
\begin{equation}
C_0 = \frac{g_0^2(\textbf{r}_m)}{\kappa\Gamma}  = \frac{3\lambda^2c}{4\pi \kappa V_{\psi}},
\label{eq:coop}
\end{equation}
is larger than one.

We consider two contributions to the cavity loss rate $\kappa$. The first is transmission of light through the cavity mirrors or absorption within them with a corresponding loss per round trip of typically $D_{abs} = 10^{-5}\leftrightarrow 10^{-3}$. The second contribution is clipping losses $D_{clip}$ coming from light of the cavity mode which misses the cavity mirrors due to their finite diameter. The cavity decay rate $\kappa$ and the corresponding cavity finesse $F$ are given by
\begin{equation}
\kappa = \frac{c}{2 L} (D_{clip} + D_{abs}) = \frac{c\pi}{L} \frac{1}{F}.
\label{eq:kappa}
\end{equation}


\subsection{Concentric cavity limit}
\label{sec:concentric}

As mentioned above, one way to enhance the cavity field at the center is to operate a spherical mirror cavity near the concentric limit. In this case, the mode volume of the fundamental Laguerre-Gaussian mode is $V_{\psi} = \frac{\pi}{4}w_0^2 L$, where $w_0$ is the waist at the focus and $L$ the cavity length, and the cooperativity $C_0$, Eq.~(\ref{eq:coop}), can be rewritten as \cite{yariv}
\begin{equation}
C_0 = \frac{3\lambda^2c}{\pi^2 \kappa w_0^2 L} = \frac{6\lambda F}{\sqrt{2RL-L^2}}
\label{eq:coop12}
\end{equation}
where $R$ is the radius of curvature of the mirrors. $C_0$ thus increases significantly for $R\rightarrow L/2$, the concentric limit. (Note that Eq.~(\ref{eq:coop12}) is valid in the paraxial approximation and breaks down when the waist approaches the wavelength.) The strong coupling rate scales as $g_0 \sim 1/w_0$. However, as the waist is reduced, the spot size on the mirror (in the far field, $L/2 \gg$ Rayleigh length) scales with $1/w_0$ and thus linearly with $g_0$. Hence, the clipping losses increase dramatically for stronger coupling.

\begin{figure}[!tbp]
  \centering
  \subfloat[]{\includegraphics[width=0.49\textwidth]{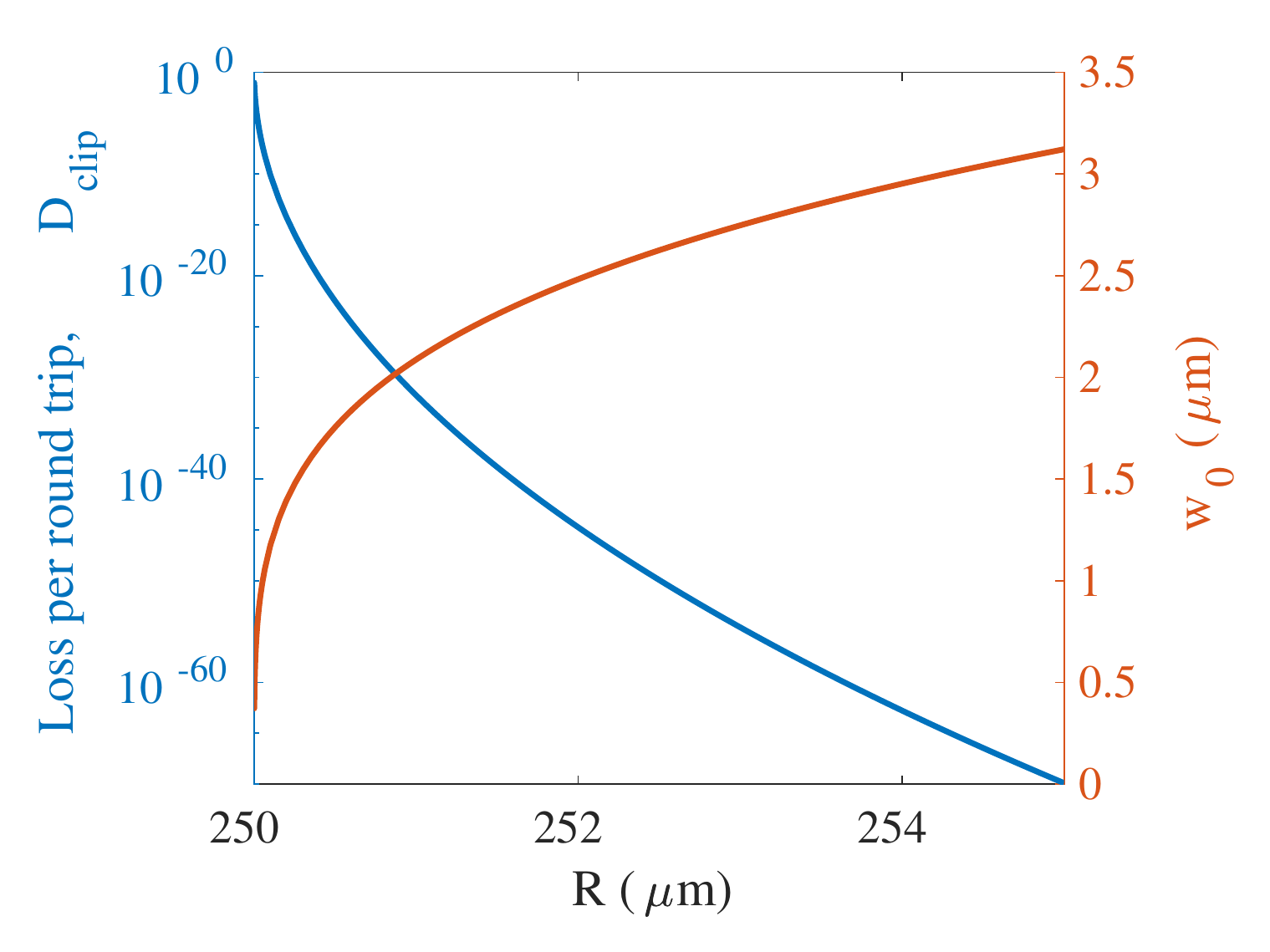}}
  \hfill
  \subfloat[]{\includegraphics[width=0.49\textwidth]{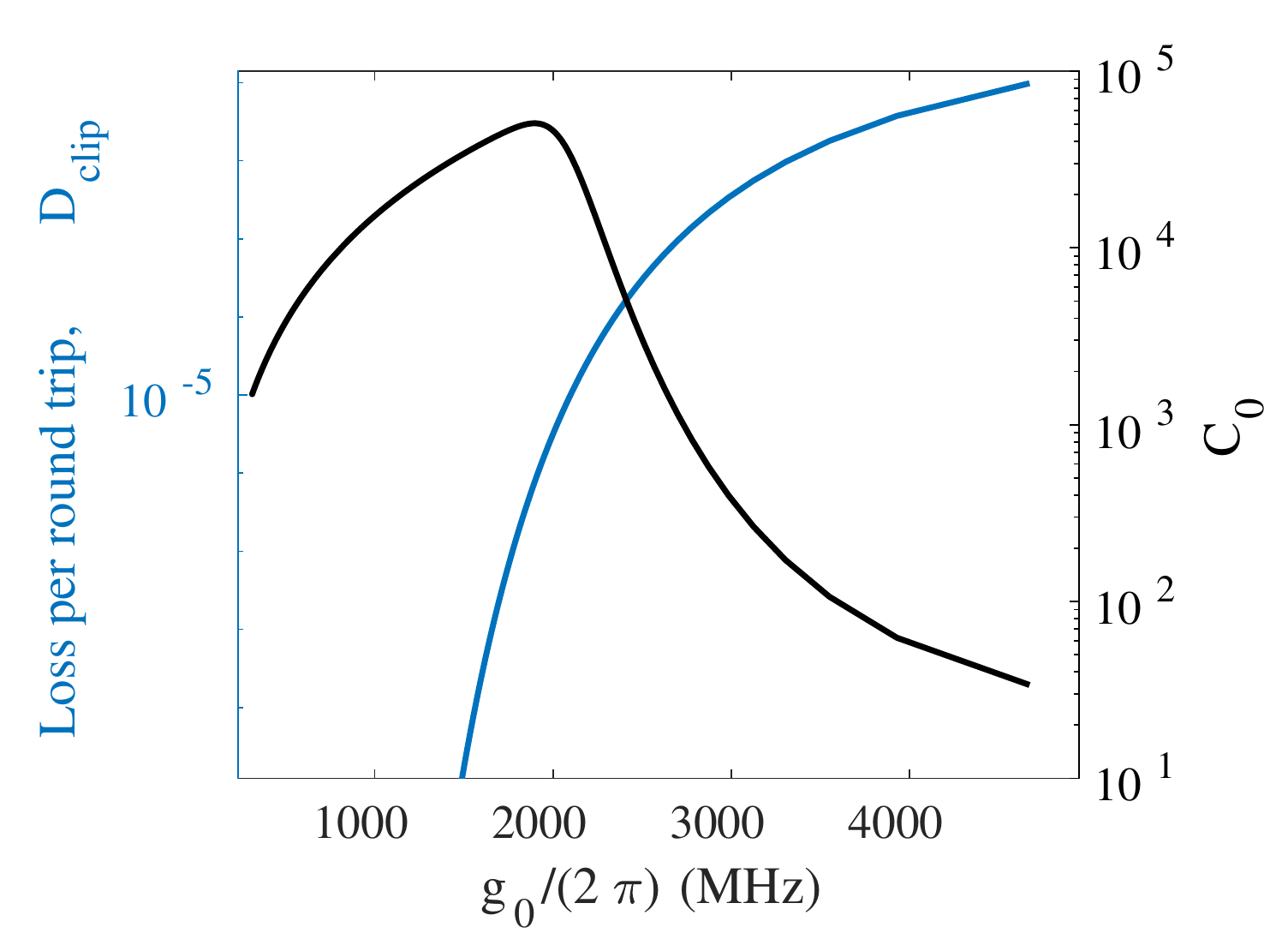}}
  \caption{(a) Clipping losses $D_{clip}$ and focal waist $w_0$ versus mirror radius of curvature for a spherical cavity near the concentric limit. Cavity parameters are $L = 500~\mu$m and mirror diameter 200 $\mu$m. (b) $D_{clip}$ and cooperativity $C_0$ corresponding to (a) plotted versus the coupling rate $g_0$ for a Ca$^+$ ion with $\Gamma_{Ca^+} = 2\pi\times 22$ MHz, $D_{abs} = 10^{-5}$.}
\label{fig:intro}
\end{figure}

Figure \ref{fig:intro} shows a numerical example of the scaling of the cavity parameters in the concentric limit. Figure \ref{fig:intro}(a) plots the loss per round trip (blue curve)  and Gaussian beam waist (red curve) as a function of radius of curvature for the fundamental cavity mode. As $R\rightarrow L/2=250\,\mu$m, the waist decreases in the center but simultaneously the mode on the mirror increases and thus clipping losses increase due to the finite mirror diameter. In Fig.\ \ref{fig:intro} (b) we re-plot the clipping loss versus the coupling rate $g_0$ (proportional to $1/w_0$ from figure (a)) which shows the increasing loss with increasing $g_0$. The black curve shows the corresponding cooperativity $C_0$: it first increases as $g_0$ increases but then rapidly falls when the spot size on the mirror exceeds the mirror diameter, thus exhibiting a clear maximum of cooperativity that can be achieved with spherical mirror cavities.


\subsection{Strong coupling rate and cooperativity enhancement by optimized cavity designs}\label{sec:gsec}

The goal of this paper is to overcome the limits found in Fig.\ \ref{fig:intro} for the cooperativity by allowing for non-spherical mirrors and using an evolutionary algorithm to optimize the shape of the mirrors. We thereby aim to increase the coupling rate $g_0$ by local field enhancement, but without the excessive clipping losses found in the concentric limit. 


\begin{figure}[tb]
  \centering
  \includegraphics[height=6cm]{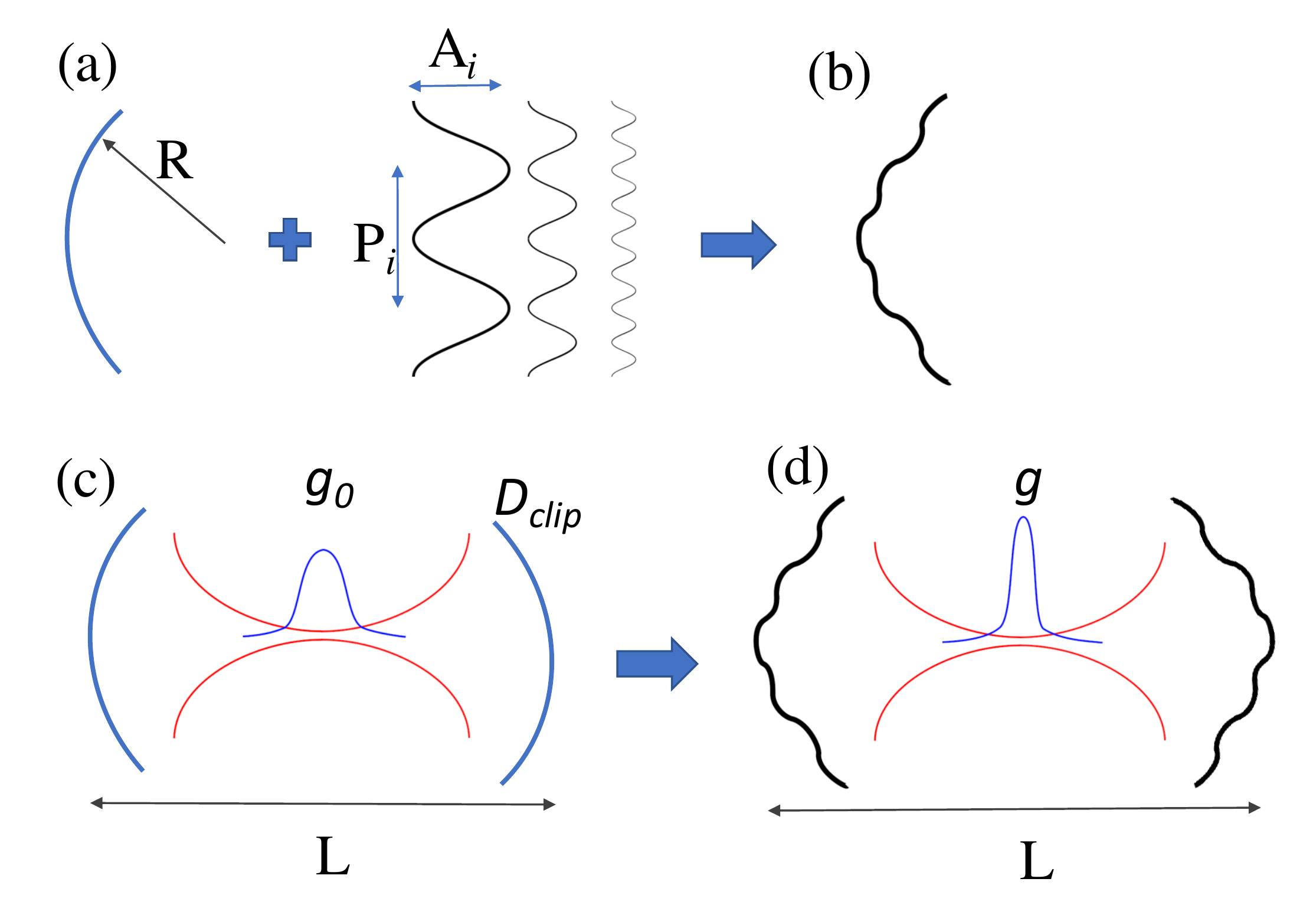}
  \caption{Schematic of our cavity optimization scheme. (a) and (b) show the mirror modification process by adding harmonic modulations to a spherical shape.  (c) and (d) indicate the cavity mode divergence by the red curves and the local field enhancement resulting from the created interference pattern by the blue Gaussian-like curves. }
  \label{fig:cavity}
\end{figure}

A schematic of this approach is shown in Fig.\ \ref{fig:cavity}. We start by choosing an appropriate spherical mirror cavity with given mirror radius of curvature and cavity length, which we use as a reference cavity. We then alter the mirror profiles, as depicted in Fig.\ \ref{fig:cavity}, and calculate the corresponding mode field $\Psi$ which we normalize to the maximum of the reference field $\psi$. Thus, $\Psi$ is the field enhancement relative to the reference cavity and we have the corresponding strong coupling rate
\begin{equation}
g(\textbf{r})=\sqrt{\frac{3\lambda^2c\Gamma}{4\pi V_{\psi} }}\Psi(\textbf{r}) = g_0 \Psi(\textbf{r}) 
\label{eq:coupl2}
\end{equation}
Likewise, the cooperativity $C$ in the center becomes  
\begin{equation}
C = \frac{g^2}{\kappa_{\Psi}\Gamma}  = \frac{3\lambda^2c}{4\pi} \frac{\Psi(\textbf{r}_m)^2}{\kappa_{\Psi} V_{\psi} } 
= C_0 \frac{\kappa}{\kappa_{\Psi}} \Psi(\textbf{r}_m)^2 
\label{eq:coop2}
\end{equation}
where $\kappa_\Psi$ is the cavity loss rate of the modified mode $\Psi$. Therefore, the cooperativity is increased by the field enhancement and/or if the cavity loss is reduced. Finally, if we assume that the reference cavity has negligible clipping losses and that mirror transmission and absorption $D_{abs}$ is the same for the reference and modified cavities, we can express the enhancement of cooperativity as
\begin{equation}
\frac{C}{C_0} =  \frac{1}{\frac{D_{clip}}{D_{abs}} + 1} \Psi(\textbf{r}_m)^2
\label{eq:coop3}
\end{equation}
where $D_{clip}$ here refers to the clipping loss of the modified cavity.


\section{Evolutionary algorithm}
\label{sec:algo}

We consider a cylindrically symmetric cavity where the mirror profile $Z(r)$ consists of a sphere with radius of curvature $R$ and some harmonic perturbations with amplitude $A_i$ and period $P_i$. We work in the paraxial limit and thus can replace the spherical profile by a parabola, leading to a modified profile described by
\begin{equation}
Z(r)=\frac{r^2}{2R} + \sum_{i} A_i \cos(r/P_i).
\label{eq:deviations}
\end{equation}
The purpose of this is to create some interference pattern and to enhance the mode field in the center $\Psi(\textbf{r}_m)$ but at the same time we want to avoid creating large clipping losses on the mirror $D_{clip}$. Figure \ref{fig:cavity} gives a schematic of the scheme.

For the numerical evaluation and optimization of our scheme we require a $Solver$ function that takes the mirror geometry as an argument and returns the cavity modes and their clipping losses, from which we select the mode with the highest cooperativity,
\begin{equation}
Z(r) \Rightarrow Solver( Z ) \Rightarrow \Psi(\textbf{r}_m,z), D_{clip} \Rightarrow C.
\label{eq:solver}
\end{equation}
The $Solver$ could be any numerical or analytic method. In this paper we implemented the solver using a mode mixing method \cite{Kleckner2010, Nina , paper1} in the paraxial approximation \cite{Lax1975}. 

\begin{figure}[tb]
  \centering
  \includegraphics[height=6cm]{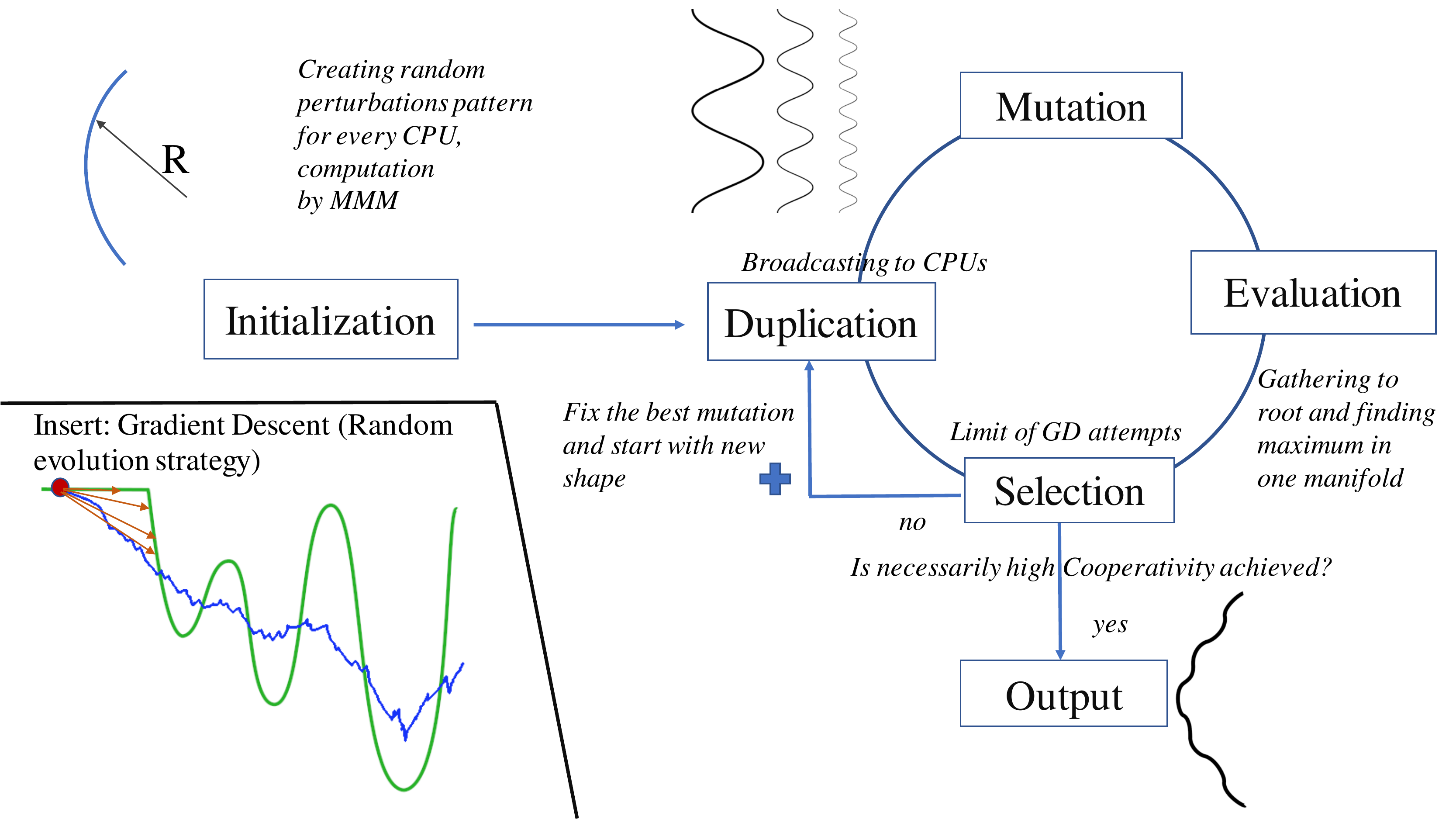}
  \caption{Schematic of the evolutionary algorithm and the gradient descent method with random evolution implementation.}
  \label{fig:algo}
\end{figure}

A schematic of our algorithm is presented in Fig.\ \ref{fig:algo}. We start with one mutation, i.e., one harmonic modulation of the mirror, represented by a couple of parameters $\theta = (A, P)$. Several instances of this geometry are then created and every instance is modified by adding a small, individual perturbation $\epsilon = (\delta A, \delta P)$ which is drawn randomly from a 2-dimensional normal distribution $N(0,\sigma_A,\sigma_P)$ with zero mean and standard deviations $\sigma_A$, $\sigma_P$. For every instance, the $Solver$ is then applied to find the cooperativity $C$ for the geometry $(\theta + \epsilon)$. The results are then combined for a gradient descent method (GD) where we compute the derivative by a probability approach used in variations analysis \cite{gd1,gd2},
\begin{equation}
\nabla C(\theta) \approx \frac{1}{\sigma_A^2 \sigma_P^2} \mathbb{E}_{\epsilon \sim N(0,\sigma_A,\sigma_P)}[ \epsilon C(\theta + \epsilon) ],
\label{eq:deriv}
\end{equation}
where $\mathbb{E}$ is the expectation value over the normal distribution. Once the GD is converged we fix this mutation, i.e., we fix the harmonic modulation $(A_1,P_1)$, and add a second perturbation $(A_2,P_2)$ which is then again optimized by GD as before. This continues until satisfactory performance of the cavity is found.

The random evolution GD is graphically shown in the insert of Fig.\ \ref{fig:algo}. The algorithm introduces some stochasticity to the results but adds significant advantages, such as the possibility to move through flat areas (when the classical gradient is zero) and to go through walls and local minima. The algorithm is also easily implemented for parallel computing with an effectiveness that grows almost linearly with computational resources \cite{gd1,gd2}. Our computer program was written in Python using the libraries numpy for computation and mpi4py for distribution over a multicore processor. 


\section{Results and discussion}\label{sec:results}

\begin{figure}[!tbp]
  \centering
  \subfloat[]{\includegraphics[width=0.5\textwidth]{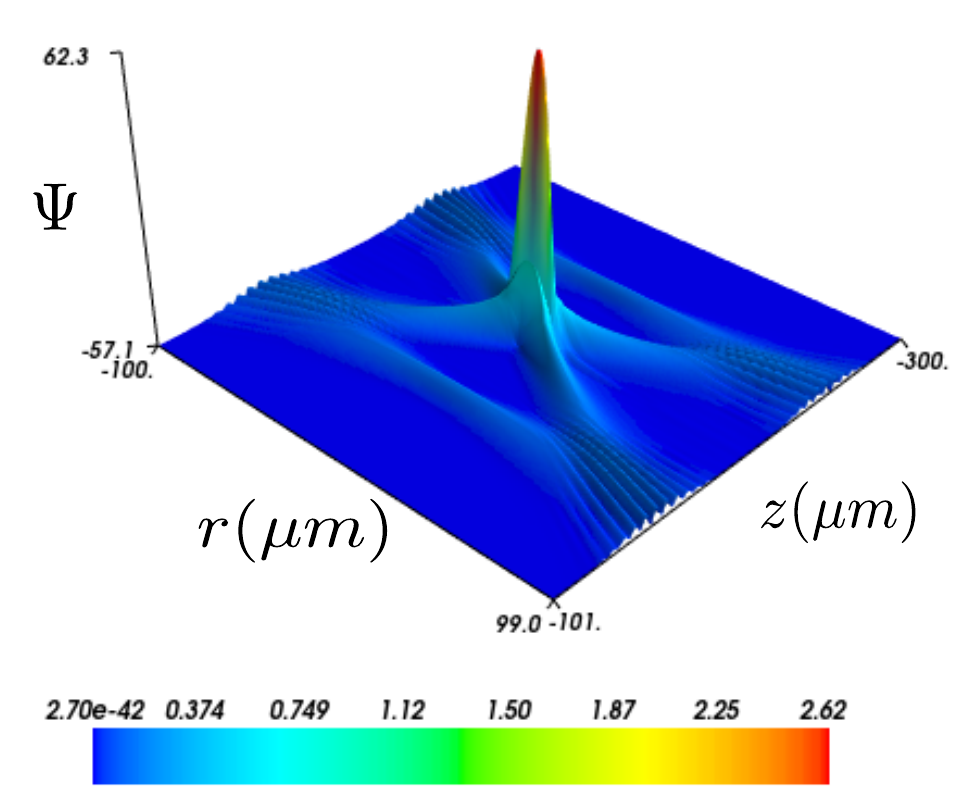}}
  \hfill
  \subfloat[]{\includegraphics[width=0.5\textwidth]{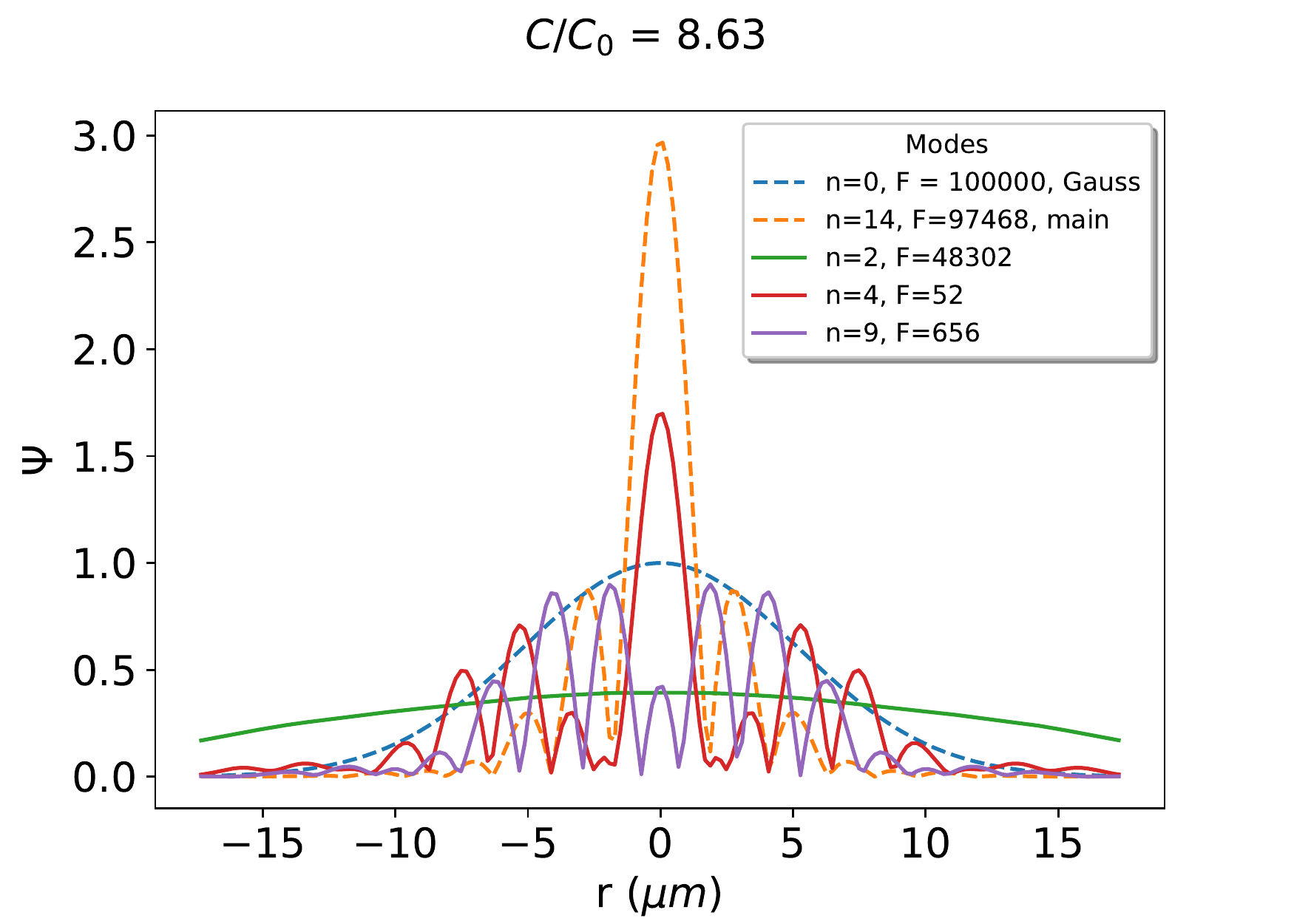}}
  \caption{(a) High cooperativity mode field $\Psi(r,z)$ for mirrors optimized by the evolutionary algorithm.  (b) Cross sections of several eigenmodes of the same cavity, with corresponding cavity finesse $F$ given in the legend. The dashed curve corresponds to the fundamental mode of spherical reference cavity with parameters $L = 500~\mu$m, $R = 400~\mu$m.}
\label{fig:modes}
\end{figure}

We present an example of our results in Fig.\ \ref{fig:modes}. Here a mirror surface with 5 harmonic mutations has been found by optimization with the evolutionary algorithm described in Sec.\ \ref{sec:algo}. The spherical reference cavity has a length of $L = 500~\mu$m, mirror radius of curvature $R = 400~\mu$m, and the operating wavelength is set to $\lambda = 0.866 ~\mu$m. The algorithm made mutations of the spherical shape with harmonic amplitudes and periods in the ranges $A_i = 0.1-0.5 ~\mu$m and $P_i = 10-30 ~\mu$m, respectively. 

The optimized eigenmode $\Psi$ with the highest cooperativity mode is shown in Fig.\ \ref{fig:modes}(a) and the cross section in the center by the orange curve in Fig.\ \ref{fig:modes}(b). For comparison, the fundamental Laguerre-Gaussian mode of the spherical reference cavity is depicted by the blue dashed line. The figure clearly shows the field enhancement achieved in the center. At the same time, for the optimized mode the clipping loss has not increased significantly and thus the cooperativity, Eq.~(\ref{eq:coop3}), has been increased by a factor of 8.63. The cavity with the optimized mirror shapes also supports other eigenmodes, some of which are shown in \ref{fig:modes}(b);  these modes have larger clipping losses and thus lower finesse, see the legend of the figure. Note that there is no unique way of defining the order of eigenmodes in such a complex mirror geometry and thus the eigenmode index $n$ is arbitrarily determined by the $Solver$.

We can clearly see that the designed superposition of harmonic perturbations on the spherical mirror shapes provides a cavity eigenmode with significant strong coupling enhancement but moderate low losses. This is a valuable enhancement for quantum optics and quantum engineering applications; a specific example will be discussed in more detail in Sec.~\ref{sec:lambda}.

We want to comment briefly on the convergence properties of our chosen algorithm. In standard GD methods depending on the choice of initial parameters the algorithm often converges to a local minimum instead of an overall, global optimum. Here we eliminate this issue by two approaches. First, we produce a superposition of mutations. Thus, if in one mutation the algorithm gets stuck in a local minimum, this is mitigated by another starting point in the next mutation. Second, we compute the gradient using the probabilistic derivative of Eq.~(\ref{eq:deriv}), which allows the algorithm to move through flat areas and through walls, see insert in Fig.~\ref{fig:algo} and references \cite{gd1,gd2}.

\begin{figure}[tb]
  \centering
  \includegraphics[height=6cm]{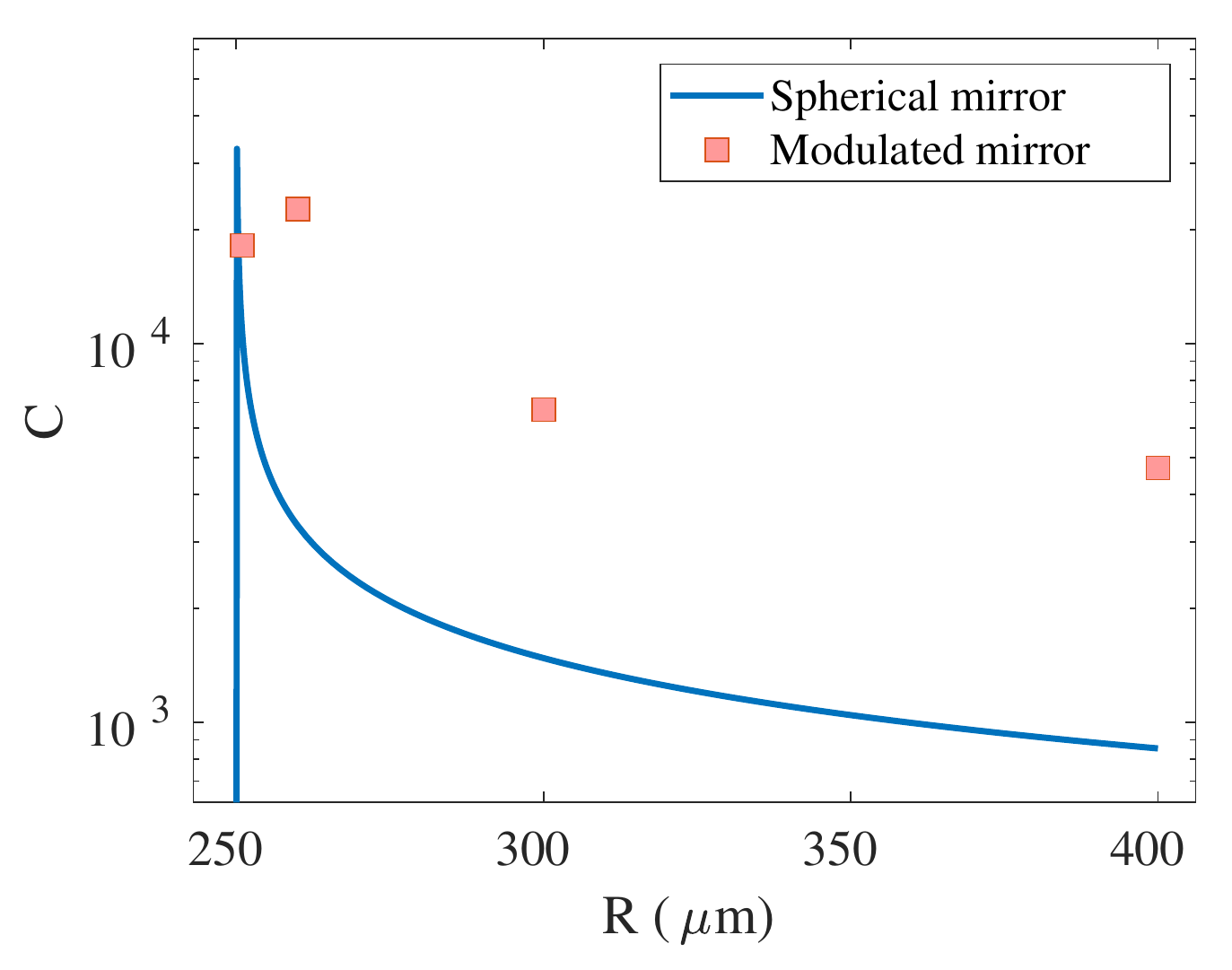}
  \caption{Comparison of cooperativities $C$ achieved by our evolutionary algorithm (red squares) with those of spherical cavities (blue curve). $D_{abs} = 10^{-5}$, $L = 500~ \mu$m.}
  \label{fig:compare}
\end{figure}

In Fig.\ \ref{fig:modes} we demonstrated an enhancement factor of 8.63 for the cooperativity relative to a spherical reference cavity deep in the stable cavity regime, i.e., far away from the unstable concentric limit $R=L/2$. This optimization was easily achieved on an 8-core processor, where the algorithm made 5 mutations and fewer than 20 iterations in every GD procedure. However, the efficiency of the algorithm and the achievable cooperativity enhancement depend on how close the reference cavity is to the concentric geometry. In Fig.\ \ref{fig:compare} we therefore show more numerical results of cooperativity enhancement achieved with our method by modulated spherical mirrors (red squares) compared to spherical cavities approaching the concentric limit (blue curve). We see that our approach provides significantly higher cooperativity over a very wide range of parameters. A spherical cavity would have to be operated in the highly unstable region with radius of curvature $R-L/2<0.05\,\mu$m to achieve comparable results.


\section{Application to quantum state transfer}\label{sec:lambda}

\subsection{Coupling to a fiber mode}\label{sec:coupling}

The same evolutionary algorithm used above for the enhancement of cavity cooperativity can be applied for the optimization of any parameter. As an example relevant for quantum information processing, we will investigate the state transfer of an ion in $\Lambda$-configuration to a single photon in an optical fiber.

\begin{figure}[!tbp]
  \centering
  \subfloat[]{\includegraphics[width=0.65\textwidth]{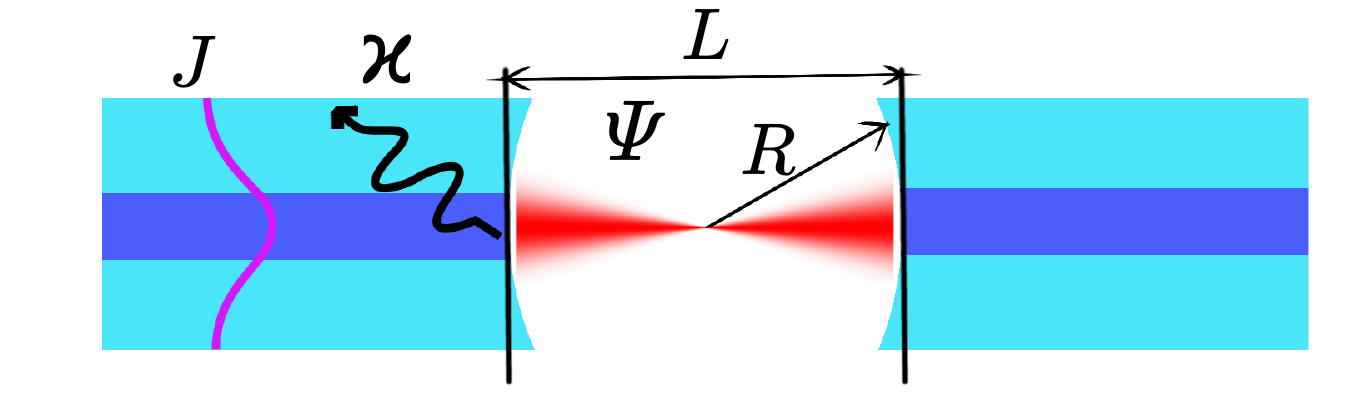}}
  \hfill
  \subfloat[]{\includegraphics[width=0.3\textwidth]{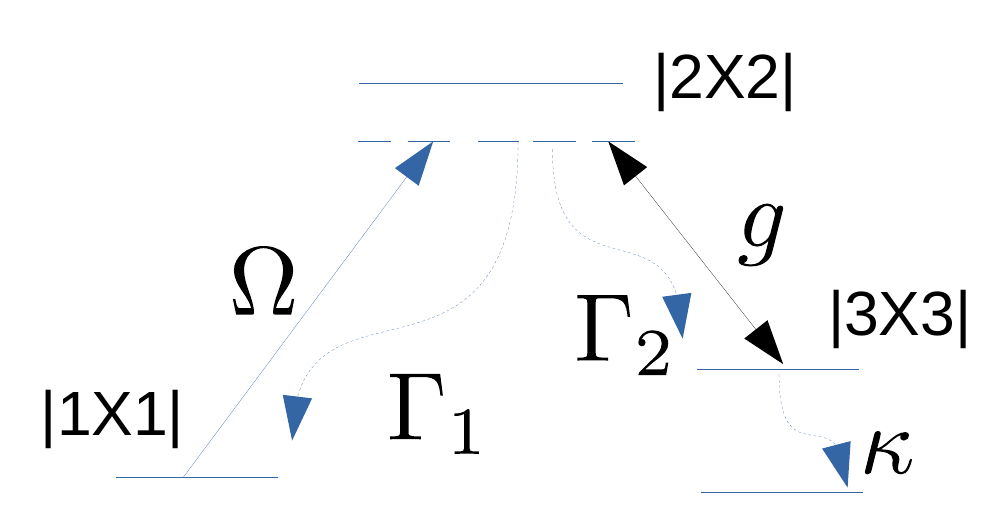}}
  \caption{(a) Coupling of a mode from a fiber tip optical cavity to an optical fiber mode. $J$ (pink curve) represents the fiber mode, $\Psi$ (red) the cavity mode, $\kappa$ the cavity loss rate, $L$ and $R$ the cavity length and mirror radius of curvature, respectively. (b) $\Lambda$-level scheme of the ion under investigation.}
  \label{fig:coupling}
\end{figure}

The ion is trapped inside a fiber tip cavity \cite{KellerFab, Hunger2010, Benedikter2015}. High cooperativity is required to transfer the ion state to a cavity photon, but good mode matching between the cavity mode through the mirror into the optical fiber mode is also necessary to optimize the outcoupling from the cavity, see Fig.\ \ref{fig:coupling}(a). Thus, the spot size as well as the curvature of the wave front of the cavity mode at the mirror should match those of the optical fiber. To determine the mode overlap we calculate the mode matching integral $Q$ between the fundamental mode $\Psi(r)$ of the cavity and the output field $J(r)$ of a single-mode fiber, 
\begin{equation}
Q = \frac{
	\left|\int \Psi(r)J^*(r) \,r dr \right|^2
}{
	\int |\Psi(r)|^2 \,r dr
	\int |J(r)|^2 \,r dr
}
\label{eq:Q}
\end{equation}
where the integration over the mirror surface is changed to an integration along the line at $z= L/2$ from 0 to the edge of the fiber in the paraxial approximation. Note that for spherical cavities $Q$ can approach 1 with the help of mode matching optics \cite{KellerFab}, but to maximize coupling from the ion to the cavity and from the cavity to the fiber simultaneously we are interested in optimizing a combination of cooperativty $C$ and overlap $Q$, which can be done by the same evolutionary algorithm presented in Sec.\ \ref{sec:algo}. Below we demonstrate this explicitly for a $\Lambda$-scheme with a calcium ion in a fiber cavity.


\subsection{ $\Lambda$-scheme}\label{sec:quant}

We consider a three level Ca$^+$ ion \cite{CaIon} in a $\Lambda$-configuration as shown in Fig.\ \ref{fig:coupling}(b). This configuration is promising for high-fidelity quantum computing and quantum communication \cite{ion1,ion2,ion3,ion4}. The system consists of two ground levels $\ket{1}, \ket{3}$ and an excited level $\ket{2}$. Raman transitions are induced by a coherent pump on the transition $\ket{1}-\ket{2}$ and the cavity mode which couples to the transition $\ket{3}-\ket{2}$. The Hamiltonian describing the coherent system dynamics is given by
\begin{equation}
H =  \Delta \ket{2}\bra{2} - \Delta \ket{3}\bra{3} +ig(a^+\sigma_3-a\sigma_3^+)+i\Omega(\sigma_1-\sigma_1^+)
\label{eq:ham}
\end{equation}
where $\sigma_3 = \ket{2}\bra{3}, \sigma_1 = \ket{2}\bra{1}$ are the transitions operators, $a$ is the cavity photon operator, $\Delta$, $g$ and $\Omega$ are detuning, strong coupling rate and coherent pumping, respectively. Incoherent decays are given by the dissipation operators%
\begin{equation}
\nu =\{ \sqrt{\Gamma_1}\sigma_1, \sqrt{\Gamma_2}\sigma_3, \sqrt{\kappa}a\}
\label{eq:dis}
\end{equation}
where $\Gamma_1$, $\Gamma_2$ and $\kappa$ are spontaneous decay rates of $\ket{3}$ to $\ket{1}$, $\ket{2}$ and cavity linewidth,  respectively. We thus solve the master equation for the density matrix $\rho$,
\begin{equation}
\frac{d\rho}{dt} = -\frac{i}{\hbar} \big[ H, \rho \big] + \sum_\nu L(\rho, \nu )
\label{eq:master}
\end{equation}
where $L$ is Lindblad superoperator given by
\begin{equation}
L(\rho, \nu ) =  \nu  \rho \nu ^+- \frac{1}{2}\big( \rho \nu^+ \nu  + \nu ^+ \nu  \rho \big).
\label{eq:limbd}
\end{equation}

For this $\Lambda$-system the expression for the cavity cooperativity needs to be corrected to 
\begin{equation}
C  = \frac{\Gamma_2}{\Gamma_1} \frac{3\lambda^2c}{4\pi} \frac{\Psi(\textbf{r}_m)^2}{\kappa V_{\Psi} }
\label{eq:coop4}
\end{equation}
since we cannot eliminate $\Gamma$ in eq.\ (\ref{eq:coop2}) because of the two different decay channels of the excited state $\ket{2}$. The branching ratio for Ca$^+$ is $\Gamma_2 / \Gamma_1 \approx 0.1$, thus reducing the cooperativity by a factor 10, %
\begin{equation}
\frac{C}{C_0} =  \frac{1}{\frac{D_{clip}}{D_{abs}} + 1} \frac{\Gamma_2}{\Gamma_1} \Psi(\textbf{r}_m)^2.
\label{eq:coop5}
\end{equation}

The efficiency of converting the ion ground state into a fiber photon is given by a combination of photon emission probability into the cavity mode $P_{e}$ and mode coupling efficiency $Q$, Eq.\ (\ref{eq:Q}). High efficiency is possible in the bad cavity regime, $\kappa \gg g^2/\kappa \gg \Gamma$, if the losses $D_{abs}$ provide coupling of the cavity mode into the fiber mode and are larger than the undesired clipping losses $D_{clip}$. In this case and for sufficiently large time, the photon emission probability is $P_{e} = \frac{C}{1+C}$ and the total efficiency $O$ of generating a fiber photon becomes \cite{kuhn}:
\begin{equation}
O  = P_{e} Q =  \frac{C}{1+C} Q.
\label{eq:effect}
\end{equation}
We can therefore use the evolutionary algorithm from Sec.\ \ref{sec:algo} to design cavity modes for an optimized total efficiency $O$. We will then use the full quantum model, Eqs.\ (\ref{eq:ham})-(\ref{eq:limbd}), to check and confirm our results.

\begin{figure}[!tbp]
  \centering
  \subfloat[]{\includegraphics[width=0.5\textwidth]{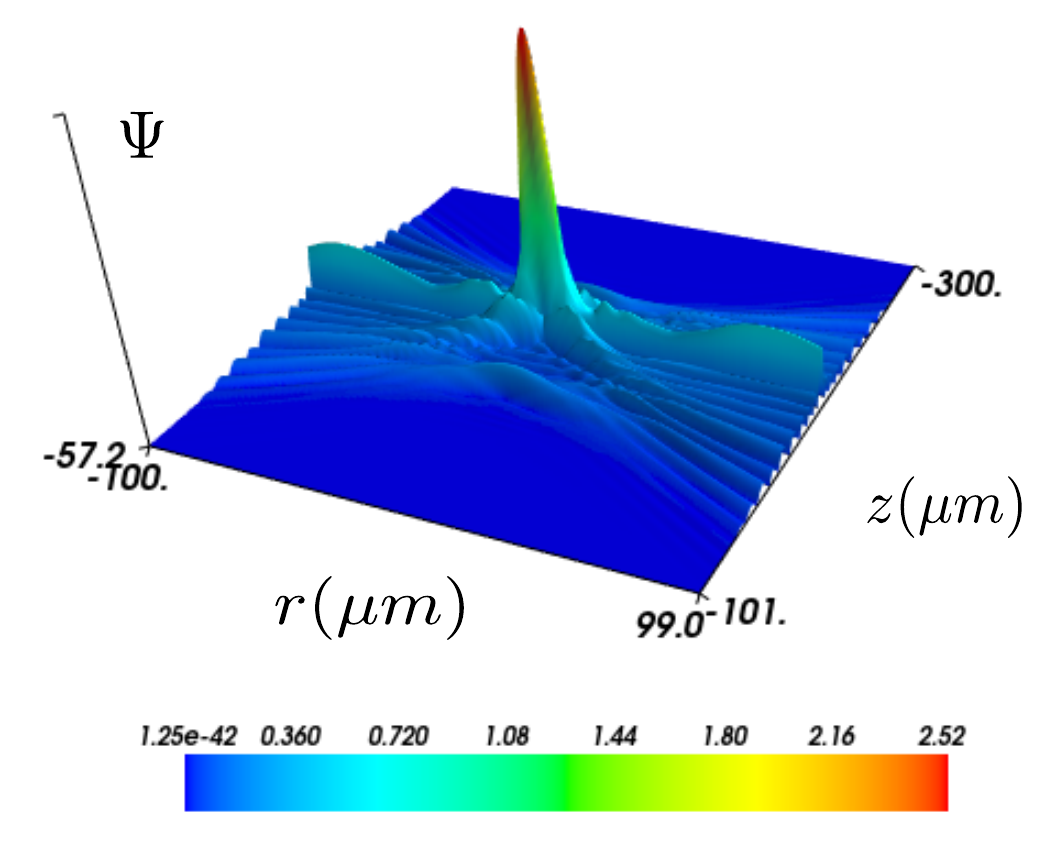}}
  \hfill
  \subfloat[]{\includegraphics[width=0.5\textwidth]{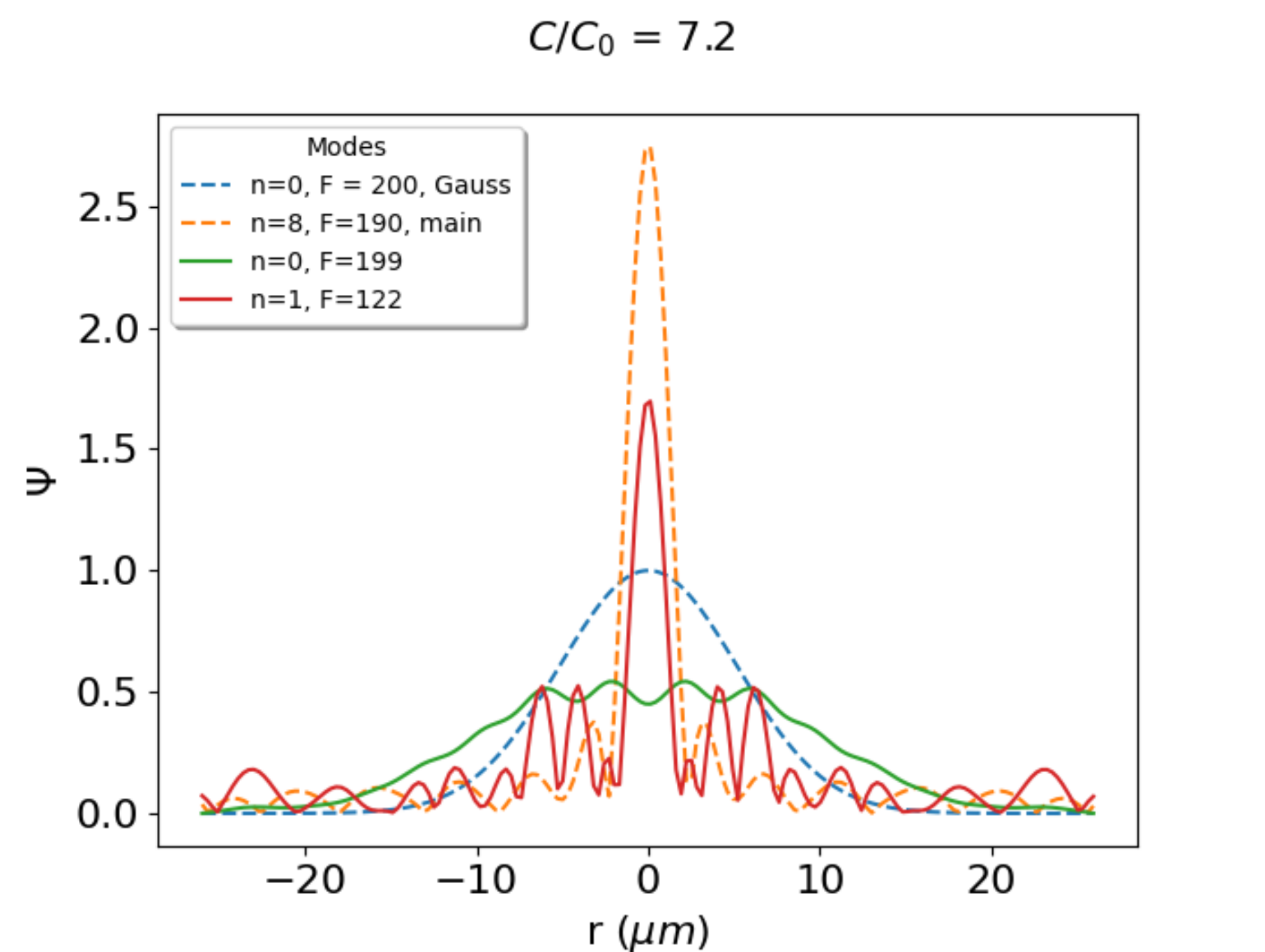}}
  \caption{(a) Optimized cavity mode field $\Psi(r,z)$ and (b) various cavity mode cross sections, similar to Fig.\ \ref{fig:modes} but optimized for maximum overall ion to fiber state transfer probability $O$. Parameters as in Fig.\ \ref{fig:modes}. }
\label{fig:qmodes}
\end{figure}

As discussed above, this approach is most valuable when we are restricted by geometry and are not able to move mirrors close to each other for a more concentric configuration. We start our optimization again with a spherical reference cavity with the following parameters: $L = 500~\mu$m, $R = 400~\mu$m, $D_{abs} = 5\cdot 10^{-3}$, $\Gamma_1 = 22\cdot 10^{6}$ s$^{-1}$, $\Gamma_2 / \Gamma_1 \approx 0.1$. This represents a fairly typical fibre-tip cavity coupled to a Ca$^+$ ion in the bad cavity limit. Assuming optimum fiber mode matching, $Q_0 \approx 1$, the cooperativity of this system is $C_0 \approx 0.17$ and thus the ion-to-fiber coupling efficiency, Eq.\ (\ref{eq:effect}), is $O_0=0.145$.

Using our evolutionary algorithm we can find a modification of the  mirror shape to achieve cooperativity and total efficiency enhancement, see Fig.\ \ref{fig:qmodes}. Compared to the optimization of cooperativity alone, Fig.\ \ref{fig:modes}, we see that we still get field enhancement at the center of the cavity, but now the field also maintains a maximum at $r=0$ on the mirrors for coupling into the fiber. The cooperativity has increased by $C/C_0 \approx 7.2$ compared to the reference cavity. The mode matching between cavity and fiber has reduced to $Q \approx 0.386$ because the cavity field is no longer Gaussian, but the overall coupling efficiency has still increased to $O \approx 0.169$ from $O_0 \approx 0.145$.

\begin{figure}[!tbp]
  \centering
  \subfloat[]{\includegraphics[width=0.48\textwidth]{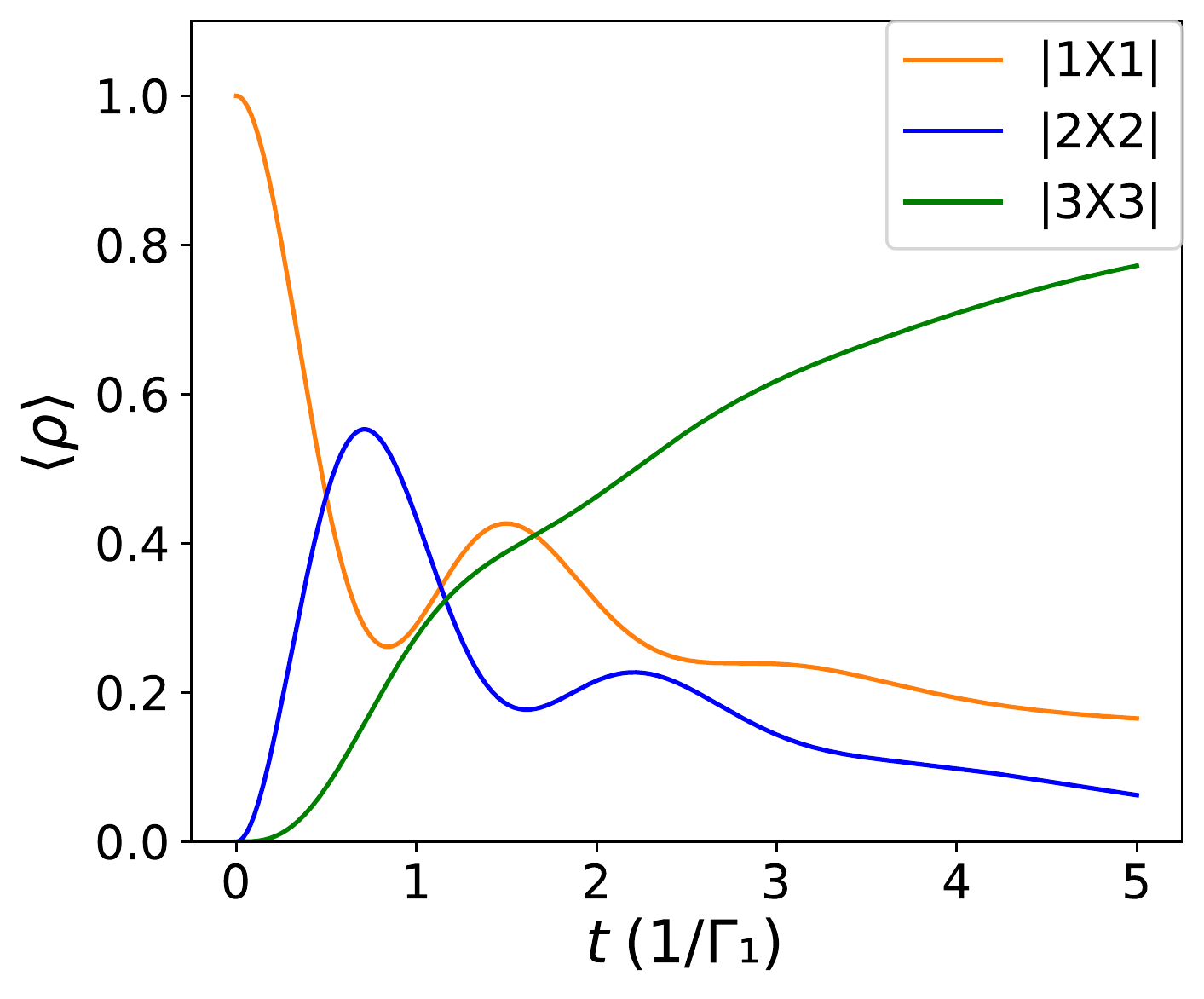}}
  \hfill
  \subfloat[]{\includegraphics[width=0.48\textwidth]{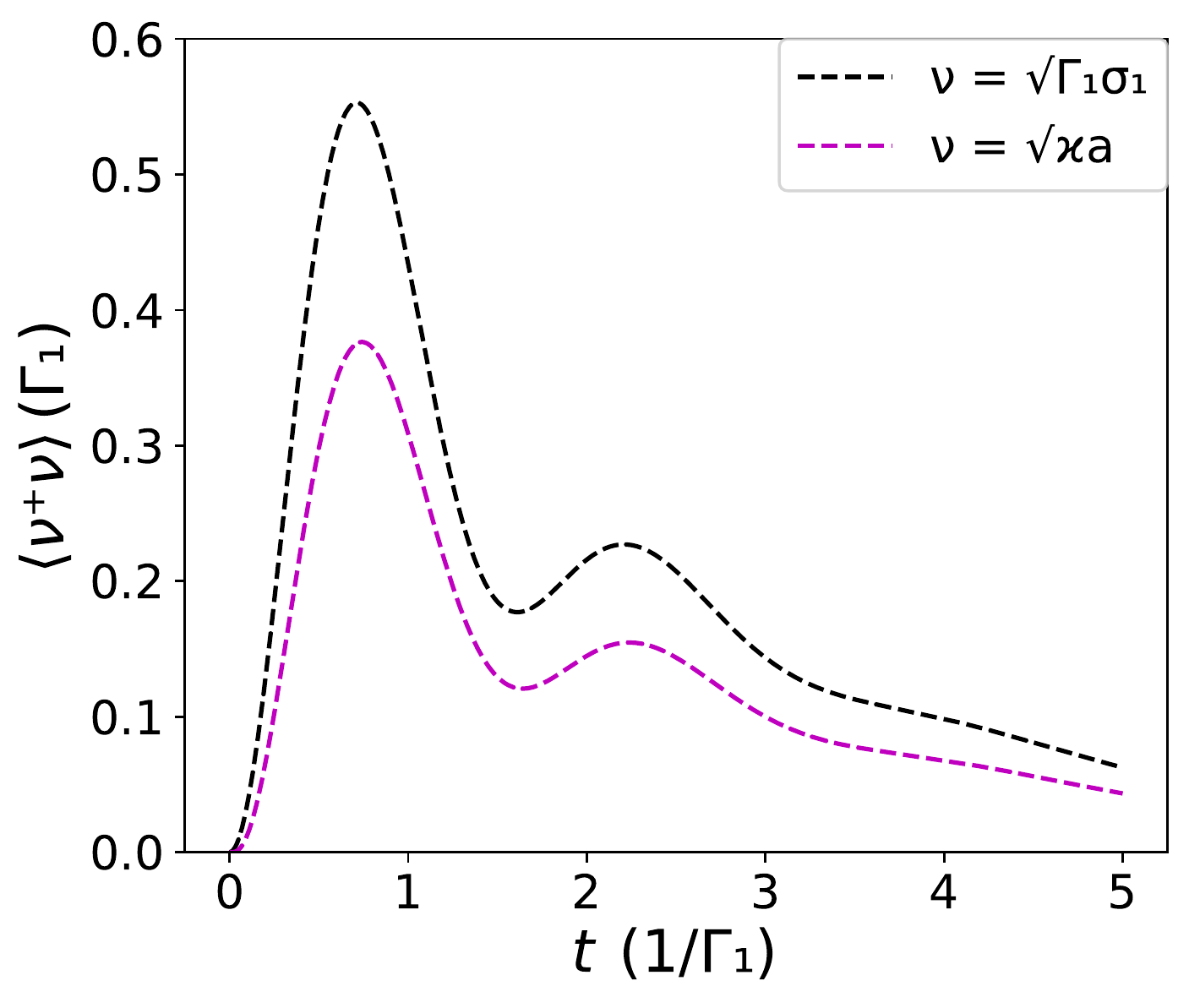}}\\
  \subfloat[]{\includegraphics[width=0.48\textwidth]{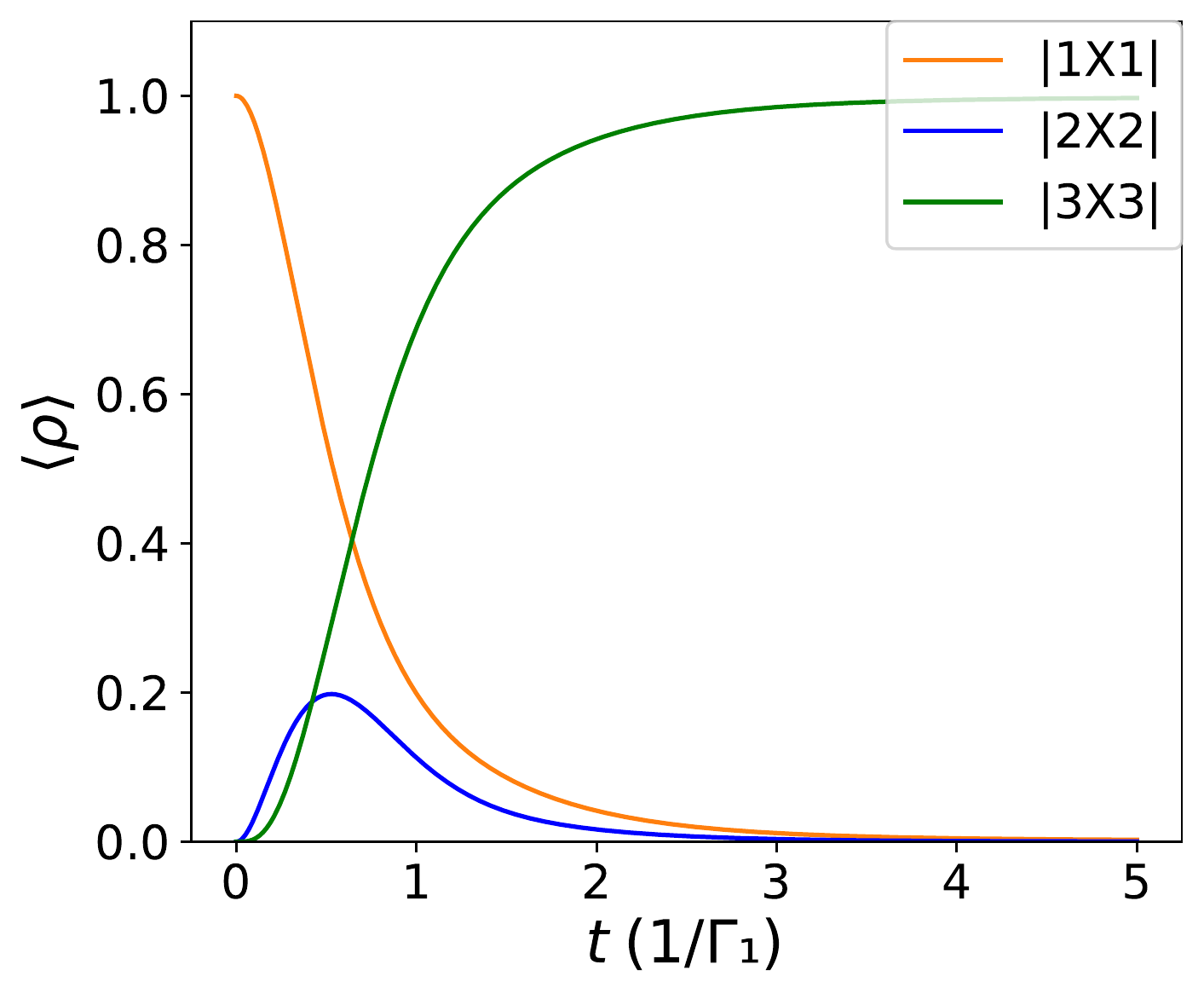}}
  \hfill
  \subfloat[]{\includegraphics[width=0.48\textwidth]{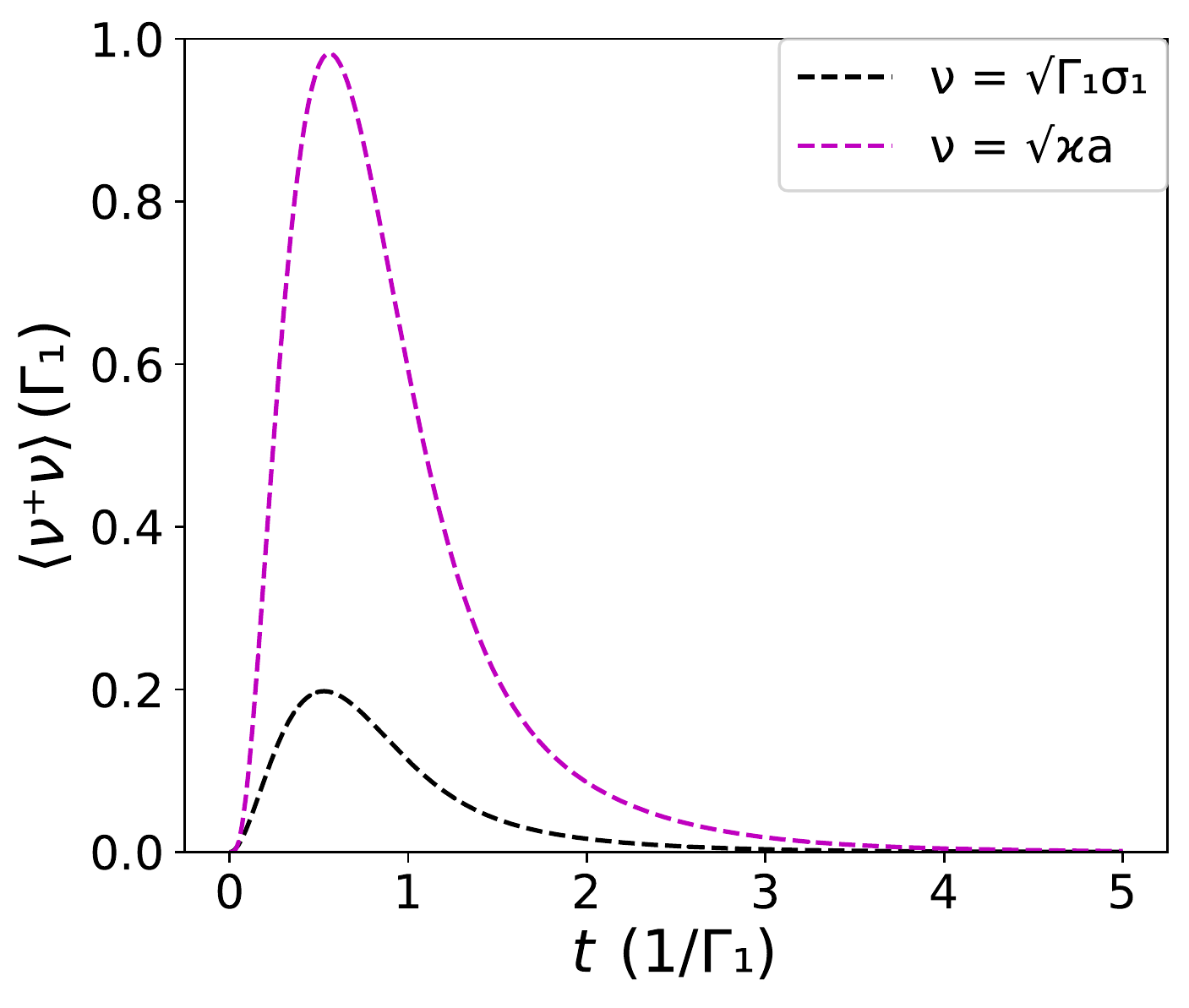}}
  \caption{Quantum dynamics of the transfer of the ion ground state to a fiber photon. (a) and (c) are the ion energy level populations for the spherical reference cavity and for the optimized cavity, respectively. (b) and (d) show photon emission and spontaneous emission probabilities corresponding to (a) and (c), respectively. Parameters are $L = 500~\mu$m, $R = 400~\mu$m, $\kappa/\Gamma_1=68.18$, $\Delta/\Gamma_1=1$, $g/\Gamma_1=3.39$ for (a,b) and 10.17 for (c,d). The pump field is $\Omega/\Gamma_1=2$.}
\label{fig:quant}
\end{figure}

The dynamics of the quantum state transfer from the ion to a fiber photon for the reference cavity is shown in Fig.\ \ref{fig:quant}(a) and (b). The detuning $\Delta$ and laser Rabi frequency  $\Omega$ are chosen to provide maximum efficiency for the fundamental Laguerre-Gaussian cavity eigenmode. The ion level populations are shown in Fig.\ \ref{fig:quant}(a): from the starting level $\ket{1}$ (orange line) we coherently excite the $\ket{1} \rightarrow \ket{2}$ transition, then via emission of a cavity photon the excited state $\ket{2}$ (blue line) is coupled to the second ground state $\ket{3}$ (green line). Fig.\ \ref{fig:quant} (b) demonstrates the corresponding ion spontaneous emission (black dashed line) and cavity decay (pink dashed line). We observe that the unwanted ion decay probability is higher than the desired cavity decay because of a large population in the excited state $\ket{2}$, Fig.\ (a), and that the overall time scale of the state transfer is rather slow because of the relatively weak ion-cavity coupling $g$, thus resulting in a  low operation frequency of the scheme.

Figures \ref{fig:quant}(c) and (d) show the equivalent quantum dynamics for the cavity mode that has been optimized with the evolutionary algorithm, Fig.\ \ref{fig:qmodes}. We note that while the total efficiency only slightly increased from 0.145 to $O \approx 0.169$, the quantum dynamics has changed dramatically compared to Figs.\ \ref{fig:quant}(c) and (d): the population of the excited state $\ket{2}$ is significantly suppressed by the faster coupling coefficient $g$, thus reducing the amount of unwanted spontaneous ion decay; the final population of  $\ket{3}$ becomes almost 1 within the chosen time scale; and the whole dynamics occurs much faster, thus enabling a much faster processing time if such a scheme is employed in a quantum information processor.

Overall, we may therefore conclude that while our optimization of $O$ has led to a reduction of the cavity to fiber mode coupling, we still gain a slight increase in the overall transfer efficiency, but even more importantly the achieved enhancement of ion-cavity coupling $g$ allows for a several times faster operation speed.


\section{Final comments and conclusions}\label{sec:conclusion}

We briefly comment here on possible experimental realizations of the mirrors designed in this paper. As we discussed in Sec.\ \ref{sec:algo}, the target mirrors are based on a spherical shape with added periodic modulations as a function of the radial coordinate $r$. We assume that these deviations are small enough to not create additional scattering losses but big enough to make fabrication possible. The typical range of perturbations predicted by our method are amplitudes of $A_i = 0.1-1 ~\mu$m and periods of $P_i = 1-100 ~\mu$m.

Such mirror machining can be achieved by many modern fabrication techniques. For example, we may shape the mirrors by laser ablation \cite{laserFab1,laserFab2,laserFab3} or focused ion beam milling \cite{fib}. Pulses of a CO$_2$ laser can be used for thermal evaporation of surface material \cite{Hunger2010, KellerFab} and laser radiation focused on the cleaved fiber ends can also produce surface qualities with extremely low roughness. Alternatively, modern micro-machining \cite{mech1, mech2, mech3} can provide hundreds of nm precision which is sufficient for our range of parameters. Sinusoidal patterns can also be achieved by laser ablation techniques using interference of laser beams to generate the required patterns.

In conclusion, we have developed an approach based on evolutionary algorithm and gradient descent methods that allows us to design optimized optical cavities for quantum optics and quantum technology applications. We demonstrated designs that achieve significant enhancement of the strong coupling rate and the cooperativity beyond the limitations of spherical, near-concentric cavities. We further demonstrated the flexibility of our numerical approach by designing a fiber-tip cavity for optimized quantum state transfer with a $\Lambda$-scheme in Ca$^+$ ions from the ion ground state to an optical fiber photon. This optimization not only led to enhancement of the transfer success rate, but also increased the operation speed of the scheme several times.

\section*{Acknowledgments}

We acknowledge financial support by the UK Quantum Technology Programme under the EPSRC Hub in Quantum Computing and Simulation (EP/T001062/1). The calculations were performed using the Iridis 5 supercomputer facilities at the University of Southampton.

\end{document}